\documentclass[hyper,letterpaper,11pt]{article}

\usepackage{euscript}
\usepackage{amssymb}
\usepackage{amsfonts}
\usepackage{amsbsy}
\usepackage{epsfig}
\usepackage{amsthm}
\usepackage{amscd}
\usepackage{amstext}
\usepackage{verbatim}
\usepackage{amsmath}
\usepackage{cancel}
\usepackage{capt-of}
\usepackage{empheq}
\usepackage{cite}

\usepackage[pdftex]{hyperref}

\def\ben{\begin{equation}}
\def\een{\end{equation}}

    \let\e=\varepsilon

\let\pa=\partial
\def\be{\begin{equation}}
\def\ee{\end{equation}}
\def\beq{\begin{equation}}
\def\eeq{\end{equation}}
\def\ba{\begin{array}}
\def\ea{\end{array}}

\def\dalemb#1#2{{\vbox{\hrule height .#2pt
       \hbox{\vrule width.#2pt height#1pt \kern#1pt
               \vrule width.#2pt}
       \hrule height.#2pt}}}

\newcommand{\bea}{\begin{eqnarray}}
\newcommand{\eea}{\end{eqnarray}}

\def\ocal{{\mathcal{O}}}

\begin{document}

\begin{center}

{ \Large {\bf
Scaling theory of the cuprate strange metals
}}

\vspace{1cm}

Sean A. Hartnoll${}^\flat$ and Andreas Karch${}^\sharp$

\vspace{1cm}

{\small
{\it ${}^\flat$ Department of Physics, Stanford University, \\
Stanford, CA 94305-4060, USA \\
${}^\sharp$ Department of Physics, University of Washington, \\
Seattle, WA 98195, USA
}}

\vspace{1.6cm}

\end{center}

\begin{abstract}

We show that the anomalous temperature scaling of five distinct transport quantities in the strange metal regime
of the cuprate superconductors can be reproduced with only two nontrivial critical exponents. The quantities are:
(i) the electrical resistivity, (ii) the Hall angle, (iii) the Hall Lorenz ratio, (iv) the magnetoresistance and (v) the thermopower. The exponents are the dynamical critical exponent $z = 4/3$ and an anomalous scaling dimension $\Phi = -2/3$ for the charge density operator.

\end{abstract}

\pagebreak
\setcounter{page}{1}

\section{Introduction}

The normal metallic state of optimally doped cuprate superconductors is highly anomalous. Many of the anomalous features -- to be reviewed shortly -- are common across different cuprate compounds and take the form of simple scaling laws. A traditional starting point for understanding the universal `strange metal' cuprate regime is to allow the electronic Green's function to obtain a singular self-energy through scattering off critical bosonic modes. A particularly successful and influential instance of this approach is the `marginal Fermi liquid' phenomenology \cite{varma}, in which a logarithmic single particle self-energy is obtained.

There is some evidence that the cuprate strange metal phase is strongly interacting and not described by quasiparticle physics. As the temperature is increased the electrical resistivity crosses the Mott-Ioffe-Regal bound \cite{Emery:1995zz, hussey, gun}. Furthermore, the low frequency (Drude-like) peaks in the optical conductivity in the strange metal phase have a width of order the temperature. See for instance data for LSCO \cite{takenaka,hussey}, Bi-2212 \cite{marel,hwang} and YBCO \cite{boris}. Such broad peaks are not consistent with the existence of long-lived quasiparticles, which should have a lifetime that is longer than the inverse temperature timescale at which generic non-conserved quantities decay \cite{Hartnoll:2014lpa}. The absence of quasiparticles will be the starting point for our characterization of strange metals.

It is a common sentiment that the ``resistivity is the first quantity to be measured and the last to be understood''. This is correct insofar as the basic explanatory unit is the single particle Green's function, from which the conductivity is derived in a potentially complicated way. However, in a strongly correlated quantum critical `soup', the basic explanatory units are operators and their scaling dimensions. The current and charge operators remain well-defined in the absence of quasiparticles, as their existence is due to Noether symmetries, while the single particle Green's functions do not. Therefore, in such circumstances, the data should be organized in terms of the currents (and hence conductivities) themselves, which will then become the first quantities to be understood.

The basic assumption we will make is that the electrical and heat current operators (and hence the associated dc conductivities and thermodynamic susceptibilities) of the cuprate strange metal phase transform covariantly under a scaling of space and time. This is tantamount to saying that the phase is quantum critical \cite{subir}, although the criticality need not necessarily originate from a quantum phase transition. For instance, it might be an intrinsically high rather than low temperature phenomenon. Putting aside for the moment the possible microscopic origin of this quantum criticality, our most important result is that nontrivial predictions can be made and verified based solely on a sufficiently sophisticated scaling analysis. In particular, we will express multiple observables in terms of only three critical exponents that we will call $\{ z,\theta, \Phi \}$. These exponents are fixed by three of the cleanest scaling laws of the strange metal: those observed in the electrical resistivity \cite{curv,phen,cooper}, the Hall angle \cite{ong,andy} and the Hall Lorenz ratio \cite{ong2}. In fact, only two of these exponents ($z$ and $\Phi$) are required to be nontrivial in order to match this data. We then go on to show that these same exponents successfully describe observed scaling in the magnetoresistance and the thermopower in the strange metal regime. Our scaling hypothesis leads to predictions for the temperature dependence of the Nernst coefficient and electronic thermal conductivity that can in principle be tested with improved data at higher (Nernst) and intermediate low (thermal conductivity) temperatures. The scaling exponents will also be shown to be consistent with the observed critical scaling of the dynamical spin susceptibility. However, because this scaling may have a different origin compared to the other quantities we discuss and our ability to reproduce it involves an additional assumption about spin, we only present these results in appendix \ref{sec:spin}.

Our analysis is phenomenological in the sense that it is not guided by any microscopic mechanism. Many scaling laws are observed in the strange metal regime, and our objective is to organize them in the most economical way possible. That said, we are directly inspired by results from model systems that are solvable using holographic duality \cite{Hartnoll:2011fn}. There, large families of quantum critical regimes without quasiparticles are found to be characterized by the three exponents that we introduce below \cite{Gouteraux:2012yr, Gath:2012pg, Gouteraux:2013oca, Gouteraux:2014hca, rrr, Karch:2014mba}. There are several obstacles, however, to our simple minded attempt to `minimally' organize the data from the cuprates. Firstly, as we discuss further below, there are multiple possible origins for scaling behavior in the cuprates. These include likely quantum critical points at both underdoping and overdoping. It is quite possible that different quantities in the same regime are controlled by different physics. Secondly, in order to implement our scaling hypothesis, we will be forced to make some kinematic assumptions about the underlying physics (see assumptions \ref{asum0} -- \ref{asum} below) that may not be correct in the final analysis. Thirdly, it is possible that only a subset of the degrees of freedom in the system are quantum critical, with the remainder described more conventionally. For instance one might have in mind hot and cold patches of a Fermi surface, or there might be multiple conduction bands. In this case the scaling contribution competes with a conventional `background' contribution and may not be dominant in all quantities. Given these caveats, we should not expect our zeroth order scaling analysis to capture every single quantity. We find it remarkable that, nonetheless, a single scaling hypothesis successfully describes the temperature dependence of many transport quantities.

\section{Critical exponents and assumptions}

The first exponent $z$ is the dynamical critical exponent and is the most familiar. This exponent parametrizes the relative scaling of space and time. In particular, in the quantum critical regime, the correlation length $\xi \sim T^{-1/z}$. We will assign units so that
\be
[k] = - [x] = 1 \,, \qquad [\omega] = - [t] = [T] = z \,.
\ee
For purposes of scaling and throughout our discussion, $\hbar = k_B = e = 1$.

The remaining two exponents $\theta$ and $\Phi$ will characterize the scaling dimensions of the entropy density $s$ and the charge density $n$, respectively. It is very important to allow these quantities to admit an `anomalous' scaling dimension, otherwise one will reach overly restrictive conclusions \cite{phillips}. The exponent $\theta$ is the hyperscaling violation exponent. It indicates the extent to which singular contributions to the entropy density do not scale like the inverse correlation volume $\xi^{-d}$, but rather $s \sim \xi^{-d+\theta}$. Here $d$ is the number of space dimensions. That is, the critical fluctuations behave as though in $d - \theta$ effective spatial dimensions. Thus we parametrize
\be\label{eq:s}
[s] = d - \theta \,, \qquad
\ee
This phenomenon is most familiar from statistical physics where it occurs in a theory above its upper critical dimension $d_\text{crit}$. In that case $d - \theta = d_\text{crit} < d$. Fermi surfaces provide a quantum scenario with nonzero $\theta = d - 1$, as the only dispersion is perpendicular to the Fermi surface \cite{hsv}. Hyperscaling violation is also ubiquitous in quantum critical phases arising in holographic theories at finite charge density \cite{Gubser:2009qt, Charmousis:2010zz, Hartnoll:2011fn, Gouteraux:2011ce,hsv}. From (\ref{eq:s}) the free energy and energy densities acquire the scalings $[f] = [\e] = z + d - \theta$. Thus we can think of quantum critical hyperscaling violation as an anomalous dimension for the energy density operator. The energy density operator couples to the system through the volume element of a background metric. This is why an anomalous dimension for the energy density is equivalent to the critical modes propagating in an anomalous number of dimensions (\ref{eq:s}). We will in fact find below that $\theta = 0$, so that hyperscaling is obeyed in our scaling analysis.

The exponent $\Phi$ is an anomalous scaling dimension for the charge density operator. It indicates that the density of charged critical fluctuations is distinct from the density of critical fluctuations contributing to the entropy. That is $n \sim s \, \xi ^{-\Phi}$ and hence we have
\be
[n] = d - \theta + \Phi \,.
\ee
For example, in a density-driven quantum phase transition, so that $\xi \sim (\mu - \mu_\star)^{-\nu}$, with $\mu$ the chemical potential, then $\Phi$ will be related to the correlation length exponent by $\nu = 1/(z - \Phi)$ \cite{jan2}. While $\Phi$ (and $\theta$) must vanish in a relativistic CFT -- this is because conserved currents saturate unitarity bounds derived from the conformal algebra -- it is allowed in more general scaling theories.\footnote{It has been argued that the density operator of a conserved charge always has its canonical dimension in any scaling theory. These arguments \cite{wen,subircharge} correctly apply when the finite temperature scaling arises from heating up a well-defined zero temperature scale invariant theory. If this assumption is relaxed, the arguments are evaded. Thus in Ref. \cite{wen} the total critical charge at a fixed time can depend on the correlation length. In Ref. \cite{subircharge}, the charge of operators in the critical theory can be allowed to depend on the correlation length. In particular, with hyperscaling obeyed, a nonzero $\Phi$ is suggestive of a temperature-dependent fraction of the bare electron going critical. This can be consistent with charge conservation in an effective critical theory. If, as will be our case, $\Phi < 0$, then the charge created by an operator grows (and diverges) as the temperature is lowered. This likely indicates an instability of the critical phase at low temperatures.} In particular, a nonzero $\Phi$ has been found necessary to understand the scaling properties of generic scaling regimes arising in compressible holographic matter \cite{Gouteraux:2012yr, Gath:2012pg, Gouteraux:2013oca, Gouteraux:2014hca, Karch:2014mba}. We will find that $\Phi$ is also essential to capture the scalings observed in the cuprate strange metal.

The dimensions of various other quantities now follow.
From the conservation laws $\dot n + \nabla \cdot j = 0$ and $\dot \e + \nabla \cdot j^Q = 0$ we obtain the dimensions of the electrical and heat current
\be
[j] = d - \theta + \Phi + z - 1 \,, \qquad [j^Q] = d - \theta + 2z-1 \,.
\ee
The heat generated by a current now implies that the electric field has dimension
\be
[E] = 1 + z - \Phi \,,
\ee
so that the chemical potential, vector potential, and magnetic field obey
\be\label{eq:B}
[\mu] = z - \Phi \,, \qquad [\vec A \, ] = 1 - \Phi \,, \qquad [B] = 2 - \Phi \,.
\ee
This scaling dimension of the magnetic field is that associated to the vector potential $\vec A$ that couples to the conserved current.
The magnetic field will also couple directly to spin, and this coupling could in general have a different scaling dimension.
Partly for this reason we focus on thermoelectric transport in the main text and only discuss spin susceptibilities in appendix \ref{sec:spin}.

From the above scaling dimensions, we can obtain the scaling dimension of the thermoelectric conductivities $\sigma, \overline \kappa, \alpha$. These are defined through the matrix
\be\label{eq:matrix}
\left(
\begin{array}{c}
j \\
j^Q
\end{array}
\right) = \left(
\begin{array}{cc}
\sigma & T \, \alpha \\
T \, \alpha & T \, \overline \kappa
\end{array}
\right) \left(
\begin{array}{c}
E \\
 - (\nabla T)/T
\end{array}
\right) \,.
\ee
The usually measured open circuit thermal conductivity is given by $\kappa = \overline \kappa - \alpha^2 T/\sigma$. In
certain non-quasiparticle circumstances these two thermal conductivities can be dramatically different \cite{Mahajan:2013cja}.
We will exclude this possibility in our assumption \ref{asum} below, and therefore $\kappa$ and $\overline \kappa$ will have the same temperature scaling.

We assume that the relevant physics occurs in $d=2$ dimensional planes. Also, we will not consider the effects of anisotropy within the planes.
The matrix of thermoelectric conductivities in (\ref{eq:matrix}) then contains six distinct in-plane observables: $\{\sigma_{xx}, \sigma_{xy}, \alpha_{xx}, \alpha_{xy}, \kappa_{xx}, \kappa_{xy}\}$. Our first objective is to express these quantities in terms of the three exponents $\{z,\theta,\Phi\}$. Certain kinematic assumptions about the emergent low energy quantum critical description of the system are necessary before we can do this. We would like to emphasize that any scaling analysis needs to take a position regarding these assumptions before it can get off the ground.

\begin{enumerate}

\item \label{asum0} The quantum critical description is assumed to be time reversal invariant. Thus the Hall conductivities must be proportional to an applied magnetic field.

\item \label{ph} The quantum critical theory is assumed not to be particle-hole symmetric. This allows the Hall conductivities divided by the magnetic field and also the thermopower to be nonzero, and to scale with temperature in a way described by the critical exponents.

\item \label{asum} The electronic and heat currents in the quantum critical theory are assumed not to overlap with any conserved or almost conserved operators. This allows the conductivities to be finite and furthermore not to be sensitive to irrelevant symmetry breaking operators (that would violate the na\"ive scaling, see e.g. \cite{Hartnoll:2012rj, Hartnoll:2014gba, Patel:2014jfa}). That is, the conductivities are described by the quantum critical scaling.

\end{enumerate}

Assumption \ref{asum0} is validated by data on the Hall conductivities, discussed below. Assumption \ref{asum} is supported by the fact that the Drude-like peaks in the optical conductivity of strange metal cuprates have widths of order $k_B T$, see e.g. \cite{takenaka, hussey, marel, hwang, boris}. In contrast, if the electrical current overlapped with an almost conserved operator, the width of the Drude peak would be the inverse lifetime of the corresponding long-lived mode, and would be narrower than $k_B T$  \cite{Hartnoll:2014lpa}. In particular, in the absence of particle-hole symmetry (assumption \ref{ph}), both electronic and heat currents will generically overlap with the momentum. Therefore, momentum must be quickly degraded in the critical theory. The system is incoherent, in the sense of \cite{Hartnoll:2014lpa}.

Assumption \ref{ph} is made because it allows us to express all quantities in terms of the three exponents $\{z,\theta,\Phi\}$ with no further input. If instead the quantum critical dynamics is particle-hole symmetric, then quantities such as the Hall conductivity and thermopower are not universal. Instead, they are sensitive to irrelevant operators that break the symmetry. Alternatively, one must deform slightly away from the quantum critical point by a small charge density to break the symmetry. This is the approach taken in e.g. \cite{Hartnoll:2007ih} and \cite{Hartnoll:2009ns}. An interesting idea to combine the observed linear resistivity with the scaling of the Hall angle in a particle-hole symmetric framework appeared recently \cite{Blake:2014yla}. Given that our
assumption \ref{ph} is primarily motivated by the simplicity of a single universal scaling analysis, such alternative particle-hole symmetric analyses are certainly worth pursuing. In fact, the canonical models of quantum criticality in finite density systems, such as the bose Hubbard model at integer filling \cite{bh}, achieve universal $xx$-component conductivities through an emergent particle hole symmetry \cite{wz,ds}. Other instances of universal conduction via emergent particle-hole symmetry occur at continuous Mott transitions in Fermi systems \cite{senmot,senmot2} and in spin density wave quantum critical points \cite{max1,max2}. In contrast, strongly interacting non particle-hole symmetric critical theories with momentum relaxation have been described in holographic settings with lattices \cite{Donos:2014uba, Donos:2014oha, Donos:2014yya}. A recent DMFT study also found scaling in the resistivity in a high temperature incoherent regime at the vicinity of a metal-insulator transition \cite{dmft}.\footnote{This DMFT study of a density-driven quantum phase transition finds $z \nu \approx 4/3$, differing from a factor of $2$ from the value $z \nu = 2/3$ that follows from our values of $z$ and $\Phi$ applied to a density-driven transistion.} In any case, our present approach is to see how far a one-parameter scaling hypothesis can take us, without reference to any particular strongly interacting microscopic model.

\section{Scaling laws}

Given the above assumptions and scaling dimensions we can conclude that, at $B = 0$
\be\label{eq:xx}
\sigma_{xx} \sim T^{(d+2\Phi-\theta-2)/z} \,,\qquad \alpha_{xx} \sim T^{(d-\theta+\Phi-2)/z} \,, \qquad
\kappa_{xx} \sim T^{(d-\theta+z-2)/z} \,.
\ee
Turning on a nonzero magnetic field we will have (to first order in the magnetic field)
\be\label{eq:xy}
\sigma_{xy} \sim B \, T^{(d+3 \Phi-\theta-4)/z} \,,\qquad \alpha_{xy} \sim B \, T^{(d-\theta + 2 \Phi - 4)/z} \,, \qquad
\kappa_{xy} \sim B \, T^{(d-\theta+z+\Phi-4)/z} \,.
\ee
There will also be magnetoresistance corrections to (\ref{eq:xx}), scaling like $B^2$, that we will discuss below.

The critical exponents can now be determined from (\ref{eq:xx}) and (\ref{eq:xy}) combined with the measurement of three quantities. The effective theory of the in-plane
transport has $d=2$ spatial dimensions. The observed linear in temperature resistivity \cite{curv,phen,cooper}
\be\label{eq:res}
\rho_{xx} \equiv \frac{1}{\sigma_{xx}} \sim T \qquad \Rightarrow \qquad 2 \Phi - \theta = - z \,.
\ee
The observed scaling of the Hall angle \cite{ong,andy}
\be\label{eq:cot}
\cot \theta_H \equiv \frac{\sigma_{xx}}{\sigma_{xy}} \sim T^2 \qquad \Rightarrow \qquad 2 - \Phi  = 2 z \,.
\ee
Finally, the observed scaling of the Hall Lorenz ratio \cite{ong2}
\be\label{eq:LH}
L_H \equiv \frac{\kappa_{xy}}{T \sigma_{xy}} \sim T \qquad \Rightarrow \qquad - 2 \Phi = z \,.
\ee
Taking the three previous equations together gives the exponents
\be\label{eq:exponents}
\fbox{
$\displaystyle
z = \frac{4}{3} \,, \qquad \theta = 0 \,, \qquad \Phi = - \frac{2}{3} \,.
$
}
\ee
It is worth emphasizing that the behavior of the Hall Lorenz ratio in (\ref{eq:LH}) requires a nonzero anomalous charge density exponent $\Phi$ within our one parameter scaling framework. The Hall Lorenz ratio is a useful observable because, unlike the usual Lorenz ratio, phonons do not contribute and so it directly probes the electronic physics. We can also emphasize that the scaling laws (\ref{eq:res}) -- (\ref{eq:LH}) have been observed in the same temperature regime of the same material: optimally doped YBCO at temperatures above the onset of superconductivity ($\sim 90\,$K) to a little above room temperature \cite{curv,ong,ong2}. The linear resistively and quadratic in temperature Hall angle have been observed in multiple cuprates.

With the exponents (\ref{eq:exponents}) at hand, we can now see whether the remaining data on strange metals is correctly reproduced. The exponents were determined from the conductivities $\{\sigma_{xx}, \sigma_{xy}, \kappa_{xy}\}$. This leaves
$\{\alpha_{xx}, \alpha_{xy}, \kappa_{xx}\}$. One can also consider the effects of deformation away from criticality by a small magnetic field $B$ (magnetoresistance).

Start with the magnetoresistance. The prediction from our scaling hypothesis is that
\be\label{eq:magneto}
\frac{\Delta \rho}{\rho} \equiv \frac{\rho_{xx}(B) - \rho_{xx}(0)}{\rho_{xx}(0)} \sim B^2 T^{(2\Phi - 4)/z} \sim \frac{B^2}{T^4} \,.
\ee
The first relation follows from the fact that the leading order magnetoresistance is expected to go like $B^2$ due to time reversal invariance, that $\Delta \rho/\rho$ is dimensionless, and the scaling of the magnetic field in (\ref{eq:B}). The second relation uses the exponents (\ref{eq:exponents}) that we found above. Remarkably, the predicted scaling (\ref{eq:magneto}) on both temperature and magnetic field is exactly what is observed for the magnetoresistance in optimally doped YBCO, LSCO \cite{magneto1} and at high enough temperatures in overdoped Tl2201 \cite{magneto2}! The magnetoresistance is in general an independent quantity than those we used to determine the exponents (\ref{eq:exponents}).\footnote{Consider the case of a particle-hole symmetric quantum critical theory. Place the theory at a small charge density, at a scale much lower than the temperature. Then $\sigma_{xx}$ is still density independent, but the Hall conductivity has to be linear in density by particle-hole symmetry. In this case one can reproduce the observed scaling of resistivity, Hall angle and Lorenz ratio by the assignment $\theta=0$, $z=2$ and $\Phi=-1$. These scaling exponents predict a magnetoresistance that goes as $B^2/T^3$ instead of the observed $B^2/T^4$, and therefore do not describe the cuprates. We see that the magnetoresistance scaling does not follow automatically from any of the other scalings, such as the Hall angle. A particle-hole symmetric theory deformed by a small charge density may be a contender to describe other strange metals.} The scaling (\ref{eq:magneto}) follows nontrivially from our scaling hypothesis and in particular the presence and value of the anomalous scaling exponent $\Phi$.

Beyond the small magnetic field limit of formulae such as (\ref{eq:magneto}), all quantities discussed in this paper are predicted to be scaling functions of the form $T^\alpha f(B/T^2)$. This will have measurable consequences at larger magnetic fields (remaining always at sufficiently high temperatures to be in the strange metal regime). From a microscopic perspective such scaling requires a velocity $v$ to construct the dimensionless quantity $(B e)/h \times [(v h)/(k_B T)]^2$. The natural microscopic velocity scale in the problem is the Fermi velocity, and indeed $v \approx v_F$ seems to be consistent with the data mentioned above.\footnote{We thank T. Senthil for this observation.} Despite the presence of a velocity scale, because $z=4/3$ the emergent quantum criticality is not characterized by linearly dispersing modes.

We now consider the thermoelectric conductivity. The critical contribution to this quantity would vanish if the critical theory were particle-hole symmetric. This conductivity is usually measured via the thermopower or Seebeck coefficient, for which our scaling predicts:
\be\label{eq:S}
S \equiv \frac{\alpha_{xx}}{\sigma_{xx}} \sim - T^{-\Phi/z} \sim - T^{1/2} \,.
\ee
We have included the minus sign for agreement with the data, the sign is not fixed by the scaling hypothesis. There is a wealth of data on thermopower in the cuprates. Often the high temperature behavior close to optimal doping is described as $S \sim - b \, T + a$, with $a,b$ constants (sometimes, with logarithmic corrections). However, the data considered in such fits is over a relatively restricted temperature range, below room temperature. Data over a larger temperature range in YBCO and LSCO clearly shows a positive, upwards curvature to the temperature dependence of $S$ at large temperatures \cite{thermo1,thermo0}. In fact, the most recent data in \cite{thermo0}, at a little above optimal doping (where the scaling is seen to cover the widest temperature range) seems to be rather well fit by $S \sim - b \, T^{1/2} + a$ over the temperature range $250 - 700 \, \text{K}$, consistent with our predicted scaling (\ref{eq:S})! We have allowed a `background' contribution described by the constant $a$. We show this fit in figure \ref{fig:thermo} below.
\begin{figure}[h]
\begin{center}
\includegraphics[height = 90mm]{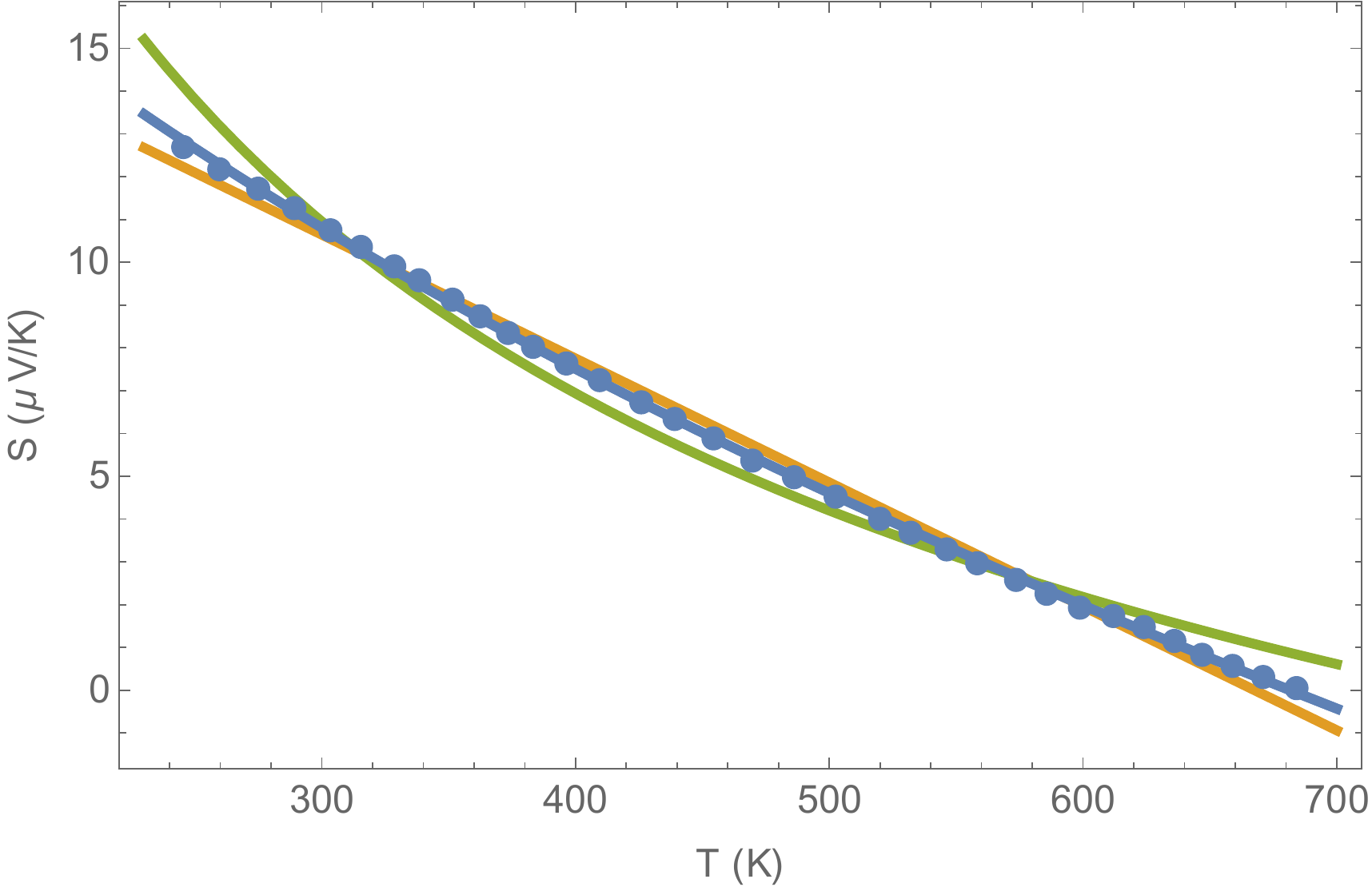}\caption{{\bf Thermopower versus temperature for LSCO} at doping $x= 0.25$, over temperatures $250 - 700K$. Dots are taken from the data curve in \cite{thermo0}. The blue line is a fit to $S \sim - b \, T^{1/2} + a$, and goes right through the data points. In contrast, fits to $S \sim - b \, T + a$ (orange line) and $S \sim b \, T^{-1/2} + a$ (green line) do not fit the data well. \label{fig:thermo}}
\end{center}
\end{figure}
In the older data, the scaling is less dramatically visible, and the exponent of the temperature dependence is consequently less robust, although still consistent with $T^{1/2}$. It would be of interest to identify a thermoelectric quantity that would show a cleaner scaling $T$-dependence over this range of temperatures, analogously to how the Hall angle cleanly reveals the scaling property of the Hall conductivity. There is also a need for high quality data over wide temperature ranges in order to sensibly discuss scaling exponents.

Consider next the Hall thermoelectric conductivity. This quantity is typically measured via the Nernst coefficient
\be\label{eq:nernst}
\nu \equiv  \frac{1}{B} \left[ \frac{\alpha_{xy}}{\sigma_{xx}} - S \tan \theta_H \right] \sim T^{-2/z} \sim T^{-3/2} \,.
\ee
We assume here that the two terms do not cancel, in which case $\nu$ would go to zero at a faster rate at large temperatures. Existing data on cuprates in the strange metal regime does indeed appear to show the Nernst coefficient going to zero at large temperatures \cite{nern1,nern2}, in qualitative agreement with (\ref{eq:nernst}). Note again that our scaling arguments say nothing about the sign of the coefficient. However, the data is not over a sufficiently large temperature regime, or of sufficiently quality, to robustly extract an exponent from the high temperature dependence. We are not aware of Nernst data over a comparable temperature range to the thermopower data we discussed in the previous paragraph. It would clearly be desirable, for our purposes, to have such data. Meanwhile, the scaling prediction (\ref{eq:nernst}) seems qualitatively reasonable, but is quantitatively neither confirmed nor excluded.

Several of the quantities we have just discussed also show interesting scaling collapse as a function of a doping as well as temperature \cite{S0, S1, H1, H2, scalingall}. This certainly strengthens the case for the existence of a quantum critical description of strange metal transport. However, understanding deformations away from the critical regime due to doping is an additional layer of subtlety that we will not attempt here. Our exponents $\{z,\theta,\Phi\}$ control the temperature dependence of the quantum critical physics itself, not deformations thereof.

The remaining transport quantity is the thermal conductivity $\kappa_{xx}$. Because phonons contribute to thermal transport, it is likely not possible to extract the electronic contribution with sufficient accuracy at the relatively high temperatures we have been focused on. If it were possible to extract the electronic contribution, the scaling prediction is that
\be\label{eq:kappa}
\kappa_{xx} \sim T^{(z - \theta)/z} \sim T \,.
\ee
If the scaling region we are studying persists down to low temperature -- which would be plausible if it is indeed due to a zero temperature quantum critical point or phase -- then by suppressing superconductivity with a large magnetic field one could hope to observe (\ref{eq:kappa}) at low temperatures. However, the observation should be made in a regime where the electrical resistivity is dominated by the linear in temperature term, not by the residual resistivity. This fact sharply distinguishes (\ref{eq:kappa}) from the standard linear in temperature thermal conductivity in low temperature regimes dominated by the residual resistivity. Existing measurements are in this residual resistivity (or indeed weak localization) regime \cite{kappa1,kappa2, kappa3}. A closely related interesting observable to look at would be thermal magnetoresistance in the strange metal normal state. This should be sensitive to purely electronic physics. The predicted scaling (at small magnetic fields) is
\be
\frac{\Delta \kappa}{\kappa} \equiv \frac{\kappa_{xx}(B) - \kappa_{xx}(0)}{\kappa_{xx}(0)} \sim B^2 T^{(2 \Phi - 4)/z} \sim \frac{B^2}{T^4} \,.
\ee
To our knowledge, measurements of this quantity do not exist in the strange metal regime.

In addition to thermoelectric transport, the spin dynamics as measured by NMR or inelastic neutron scattering is an important probe of cuprate systems. We are able to reproduce observed scaling in these quantities. However, because (i) we need to assume that the scaling dimension of the magnetic field coupling to the spin is the same as that coupling to the conserved electric current and (ii) the most dramatic scaling is observed only in underdoped LSCO and may therefore have a different origin to the scaling discussed so far, we have left this discussion to appendix \ref{sec:spin}.

We now turn to thermodynamic quantities. From scaling we obtain the specific heat and magnetic susceptiblity
\bea
c & \equiv & - T \frac{\pa^2 f}{\pa T^2} \sim T^{(d-\theta)/z} \sim T^{3/2} \,, \\
\chi&  \equiv & - \frac{\pa^2 f}{\pa B^2} \sim T^{(z-\theta+2\Phi - 2)/z} \sim T^{-3/2} \,.
\eea
 To extract the electronic specific heat experimentally, one must subtract the phonon contribution. Within a certain subtraction scheme,
 experiment on optimally doped YBCO gives $c \sim T$ from $T_c$ to room temperature \cite{cs}.
 This is of course the usual Fermi liquid scaling. Furthermore, the measurement finds that
 $c/(\chi T)$ is given by the free electron Wilson ratio over this temperature range \cite{cs}. The thermodynamic quantities therefore appear to be rather conventional and not governed by the same quantum critical dynamics as the transport quantities we considered above. This suggests the existence of degrees of freedom that contribute to and dominate thermodynamics but not transport. For instance, such degrees of freedom might be localized. We discuss various possibilities below. An important aspect of the difference between transport and thermodynamics is that we have found that transport is characterized by hyperscaling physics (i.e. $\theta = 0$), whereas Fermi surface thermodynamics requires $\theta = d-1$ \cite{hsv}. It may be possible to identify a scaling contribution on top of the dominant conventional background. A fit to $c/T \sim a - b T^{3/2}$ correctly captures the slight decrease of $c/T$ with temperature, but the power is not strongly constrained. The magnetic susceptibility has been more thoroughly characterized. To cleanly identify a temperature scaling, following our experience above, we expect to need slightly overdoped samples with data points extending to high temperatures. Data in this regime, such as \cite{mag1,mag2,mag3}, do show the susceptibility decaying at large temperatures with a positive curvature. While some of the data is not incompatible with the form $\chi \sim a + b T^{-3/2}$ at large temperatures, the fit is not compelling. If the scaling contribution is there at all it must be strongly subleading compared to a more conventional contribution.

To summarize the above: We have shown that simple kinematical assumptions plus two nontrivial critical exponents ($z$ and $\Phi$) can capture many of the observed scaling relations in the strange metal regime of the cuprates. Specifically we successfully described scalings in (i) the electrical resistivity, (ii) the Hall angle, (iii) the Hall Lorenz ratio, (iv) the magnetoresistance and (v) the thermopower. In appendix \ref{sec:spin} we also, possibly coincidentally, reproduce (vi) the dynamical spin susceptibility. These facts do not tell us what the underlying mechanism causing the scaling is, but will surely constrain it. It would be wonderful to find a compelling microscopic theory predicting the values of the exponents in (\ref{eq:exponents}).

\section{Discussion}

Quantum critical points are one possible origin of scaling laws. A key signature of a quantum critical point is a divergence in the effective quasiparticle mass as detected via quantum oscillations in large magnetic fields. Such divergences are observed in underdoped cuprates at $p \approx 0.08$ \cite{qcp1}, as well as in overdoped cuprates at $p \approx 0.18$ \cite{qcp2}. These values of carrier density are precisely those where the largest magnetic field is required to suppress superconductivity \cite{bigmag1,qcp2}. The nature of the critical points is likely quite different. The underdoped critical point appears to be associated with the onset of magnetic order and of Mott insulation. The overdoped critical point (or points) appears to be associated to the closure of the pseudogap and Fermi surface reconstruction. It is unclear a priori which critical points control which parts of the phase diagram, an issue that is further complicated by the fact that the location of the critical points may move due to the large magnetic fields and/or the superconducting condensate \cite{movingqcp}. Most of the observables we have discussed -- the exception is the dynamical spin susceptibility discussed in the appendix -- have been measured in optimally or slightly overdoped materials. The linear in temperature resistivity regime, for instance, is centered around slightly overdoped samples \cite{cooper}. Thus one possibility is that the critical exponents we have found are associated with a quantum critical point describing the closing pseudogap and/or Fermi surface reconstruction. The scaling in the spin susceptibility would then either be due to a different critical point at lower doping, or could be induced by coupling to the critical modes at the higher doping critical point.

The conventional behavior of the thermodynamic quantities $\chi$ and $c$ above is in tension with the anomalous transport and needs to be explained.
Three simple scenarios are the following: (a) localized degrees of freedom dominate thermodynamics but do not contribute directly to transport, (b) neutral degrees of freedom such as a spinon Fermi surface dominate thermodynamics but do not contribute to charge transport, (c) the critical physics is localized in momentum space, involving e.g. charge or spin density waves excitations, whereas thermodynamics is dominated by the `cold' remainder of the Fermi surface. A weakly interacting picture has long suggested that transport in such scenarios will be dominated by the conventional `cold' quasiparticles that short-circuit the critical modes \cite{Hlubina}. However, it has recently been argued that strong coupling between different patches of the Fermi surface can invert this logic, so that the critical `hot spots' can dominate transport \cite{Patel:2014jfa}.

Another possibility, not necessarily in contradiction with the quantum critical scenario, is that the scalings are a consequence of a high temperature, incoherent phase characterized by the absence of any long-lived excitations \cite{Hartnoll:2014lpa}. Most of the data we have incorporated within our scaling framework is indeed at temperatures ranging from above the optimal $T_c$ to well above room temperature. Some scaling properties in this regime may well be distinct from any scaling emerging at very low temperatures, cf. \cite{dichot}.

To end with, we comment on several measurements of temperature and frequency scaling in the cuprates that we have not yet discussed. Also, we will make some remarks on the implications of our results for the understanding of high temperature superconductivity.

A well-known measurement in Bi-2212 found the optical conductivity to scale as $\sigma(\omega) \sim \omega^{-2/3}$ over an intermediate frequency regime \cite{marel}. This observation does not fit into our current scaling analysis. However, the frequencies at which this scaling is observed correspond to temperature scales greater than about $1500$K. This is a different (higher) temperature range than the rest of the observables we have fit in this paper. It seems plausible, then, that this scaling has a different origin. At lower frequencies, we have already mentioned the fact that the optical conductivities of cuprates have a low frequency peak with width of order $\Gamma \sim T$, see e.g. \cite{hussey, takenaka, marel, hwang, boris}. This latter fact is compatible with our single-parameter scaling hypothesis.

An energy width (quasiparticle lifetime) of order $\Gamma \sim T$ is also observed around optimal doping in ARPES measurements, together with $\omega/T$ scaling \cite{arpes}. While microscopic single particle lifetimes are not natural observables from our non-quasiparticle perspective, the absence of a new scale in these lifetimes is again compatible with our single-parameter scaling hypothesis.

An important motivation to understand the strange metal regime of the cuprates is of course the hope that it will help
to explain the emergence of high temperature superconductivity from this regime. Once the quantum critical nature
of the strange metal is established, the onset of superconductivity must also be understood with this framework.
In particular, as with the current operators we have considered here, the essential feature of the `Cooper pair operator'
$\ocal$ that condenses will be its scaling dimension $\Delta_\ocal$. This quantity can in principle be measured
through the pair susceptibility \cite{ferrell,scalapino,jan}. The dimension of the Cooper pair operator appears as a natural organizing
principle in holographic studies of superconductivity \cite{Denef:2009tp, Hartnoll:2011fn}, following \cite{Gubser:2008px, Hartnoll:2008vx, Hartnoll:2008kx}. It would be interesting if, in analogy to our study here, the exponent $\Delta_\ocal$ could be shown to tie together various different experimental observables.

We have restricted attention to the cuprates in this study.
It would be natural to adapt our analysis to other well-characterized strongly interacting materials that exhibit scaling in the temperature dependence of observables \cite{Tlin}. New experiments will likely be required. The exponents need not be the same as for the cuprates, of course. Natural candidates include heavy fermions, pnictide superconductors and ruthenates.

\section*{Acknowledgements}

We would like to thank Blaise Gout\'eraux, Nigel Hussey, Aharon Kapitulnik, Dmitri Khveshchenko, Steve Kivelson, Philip Phillips, Subir Sachdev, Todadri Senthil and Jan Zaanen for helpful discussions on related topics. The work of SAH is partially supported by a US Department of Energy Early Career Award and by a Sloan fellowship. The work of AK is partially supported by the US Department of Energy under grant number DE-SC0011637.

\appendix

\section{Scaling of the spin susceptibility}
\label{sec:spin}

Spin susceptibilities are measured via the coupling of spin to a magnetic field. However, in a non-relativistic theory the scaling dimension of this coupling need not be the same as the dimension of the magnetic coupling to the electric current. Therefore, to describe the potential scaling of spin susceptibilities, we have the freedom to choose a new exponent. In this appendix we show that certain measurements of quantum critical scaling in cuprate spin susceptibilities can be reproduced if we assign the magnetic field coupling to spin the same dimension (\ref{eq:B}) as that coupling to the electric current.

Quantum critical scaling has been reported in the momentum-dependent dynamical spin susceptibility of La$_{1.86}$Sr$_{0.14}$CuO$_4$ as measured by inelastic neutron scattering \cite{aeppli}:
\be\label{eq:exp}
\lim_{\omega \to 0} \frac{\chi''(\omega,q_\star,T)}{\omega} \sim \frac{1}{\left[\Delta_q(T) \right]^{3 \pm 0.3}} \,.
\ee
Here $q_\star$ is the location in momentum space of the peak in the susceptibility and $\Delta_q$ is the temperature dependent width of the peak. The prediction from our scaling is
\be\label{eq:goodq}
\lim_{\omega \to 0} \frac{\chi''(\omega,q_\star,T)}{\omega} \sim \left[\Delta_q(T)\right]^{-\theta + 2\Phi -2} \sim
\frac{1}{\left[\Delta_q(T) \right]^{10/3}} \,.
\ee
The exponent is seen to be within the error bars of the experimental measurement (\ref{eq:exp})! The compound used for this measurement is slightly underdoped, while our emphasis in the main text was on slightly overdoped samples. It is not clear, therefore, whether it is correct to include this quantity in our analysis, as the scaling may have a different origin, as we mentioned in the discussion section. Furthermore, the scaling is not seen in YBCO, and so may be less universal than the transport results discussed above. In any case, the observed scaling in LSCO is reproduced by our exponents. The data in \cite{aeppli} was also used to argue for a $z=1$ dynamical critical exponent. However, that data does not seem to be clean enough (in particular, there is a non-negligible constant offset to the scaling of $\Delta_q$ with $T$) to exclude our preferred $z=4/3$, which is not such a different value.

The same inelastic neutron scattering data \cite{aeppli} has furthermore been used to extract the momentum-integrated susceptibility. The result is that \cite{ins}
\be\label{eq:taueff}
\tau_\text{eff} \equiv T \int d^2q  \lim_{\omega \to 0} \frac{\chi''(\omega,q,T)}{\omega} \sim \text{const.} \,,
\ee
in the quantum critical regime. Our scaling predicts
\be\label{eq:allq}
\tau_\text{eff} \sim T^{(-\theta + 2 \Phi + z)/z} \sim \text{const.} \,,
\ee
again in agreement with the data! For both (\ref{eq:goodq}) and (\ref{eq:allq}) to work, the value $z = 4/3$ was crucial.

The same quantity $\tau_\text{eff}$ in (\ref{eq:taueff}) can be expected to determine the NMR relaxation rate $1/T_1$ of nuclear spins coupled to e.g. an antiferromagnetic order parameter. Indeed, $1/T_1$ at the ${}^{63}$Cu sites in LSCO is also measured to be constant in the high temperature quantum critical regime \cite{NMR}. This is again in agreement with our scaling according to (\ref{eq:allq}).

\end{document}